\documentclass[aip,reprint,prl]{revtex4-1}

\usepackage{graphicx}
\usepackage{amsmath}
\usepackage{amsfonts}
\usepackage{url}
\usepackage{color}
\usepackage{bm}
\usepackage{bbold}
\usepackage{graphics}
\usepackage{paralist}

\begin{document}

%\title{A black-box local Gaussian representation for the simulation of correlated extended systems}
%\title{Transforming local Gaussian basis sets into plane waves for the simulation of correlated periodic systems}
\title{From plane waves to local Gaussians for the simulation of correlated periodic systems}

\author{George~H.~Booth}
\email{george.booth@kcl.ac.uk}
\affiliation{Department of Physics, King's College London, Strand, London, WC2R 2LS, UK}
\author{Theodoros~Tsatsoulis}
\affiliation{Max Planck Institute for Solid State Research, Heisenbergstra{\ss}e 1, 70569 Stuttgart, Germany}
\author{Garnet~Kin-Lic~Chan}
\affiliation{Department of Chemistry, Frick Laboratory, Princeton University, Princeton, New Jersey 08544, USA}
\author{Andreas~Gr\"uneis}
\email{a.grueneis@fkf.mpg.de}
\affiliation{Max Planck Institute for Solid State Research, Heisenbergstra{\ss}e 1, 70569 Stuttgart, Germany}

\begin{abstract}

   We present a simple, robust and black-box approach to the implementation and use of local, periodic, atom-centered Gaussian basis functions
   within a plane wave code, in a computationally efficient manner. The procedure outlined is based on the representation of the Gaussians within    a finite bandwidth by their underlying plane wave coefficients. 
The core region is handled within the projected augment wave framework,
by pseudizing the Gaussian functions within a cut-off radius around each nucleus, 
smoothing the functions so that they are faithfully represented by a plane wave basis with only moderate kinetic energy cutoff.
%% Issues to do with the representation of the non-analytic core
%%    region are resolved via the projected augmented wave framework, and a pseudization procedure which smooths the core region of the Gaussian
%%    functions 
   To mitigate the effects of the basis set superposition error and incompleteness at the mean-field level
introduced by the Gaussian basis, we also propose a hybrid approach,
   whereby the complete occupied space is first converged within a large plane wave basis, and the Gaussian basis used to construct a complementary virtual space for the application of correlated methods.
   We demonstrate that these pseudized Gaussians yield compact and systematically improvable spaces with an accuracy
   comparable to their non-pseudized Gaussian counterparts.
   A key advantage of the described method is its ability to efficiently capture and describe electronic correlation effects of weakly bound and
   low-dimensional systems, where plane waves are not sufficiently compact or able to be truncated without unphysical artifacts.
   We investigate the accuracy of the pseudized Gaussians for the water dimer interaction, neon solid and water adsorption on a LiH surface, at the level of second-order M\o ller--Plesset perturbation theory.
   
%    We outline a simple method to convert atom-centered local Gaussian basis functions into a plane wave basis set
%    such that the resultant basis functions can be employed in a computationally efficient manner using in principle any
%    electronic structure code based on the projector augmented wave method and plane wave basis sets for periodic systems.
%    The described procedure is based on pseudizing the sharply peaked radial Gaussian basis
%    sets inside the projector augmented wave spheres. The resultant radial basis functions are termed pseudized Gaussians and
%    can be expanded in a plane wave basis set using only moderate kinetic energy cutoffs.
%    We demonstrate that the pseudized Gaussians allow for the construction of compact and systematically truncatable virtual
%    orbital basis sets that can be used for correlated wavefunction based theory calculations. Furthermore we show that the obtained
%    accuracy using the pseudized Gaussians is comparable to their non-pseudized Gaussian counterparts.
%    The key advantage of the described methods as opposed to pure plane wave basis sets lies in their ability
%    to efficiently capture and describe electronic correlation effects of weakly bound molecular as well as low-dimensional systems,
%    where plane wave basis sets are not sufficiently compact or systematically truncatable.
%    We investigate the accuracy of the pseudized Gaussians for the water dimer interaction, Neon solid and water adsorption on an LiH
%    surface using second-order M\o ller-Plesset perturbation theory.
    
\end{abstract}

\maketitle

\section{Introduction}
\label{sec:intro}

In the development of first principles electronic structure methods for extended systems, a huge amount of research effort is expended exploring different function
spaces in which to expand the electronic wavefunctions. This is a critical design choice since the rate of convergence of the desired properties of the system with 
respect to the size of this
function space will substantially impact the computational cost and feasibility of calculations, and go a long way to determining the utility of the approach. 
In this paper, we will consider a simple, robust protocol for building a periodic Gaussian basis from an underlying traditional plane-wave expansion. In particular, we focus
on the ability to converge the virtual manifold of states, required for correlated, post-mean-field calculations, such as the
Random Phase Approximation (RPA), GW theory, M\o ller-Plesset theory (MP2), coupled-cluster theory, as well as multi-configurational strong correlation approaches\cite{Helgie}. 
The computational cost of these methods grows between quadratically and exponentially with respect to the number of virtual states, and therefore the ability 
to span the relevant parts of this space with as few functions as possible becomes a critical design decision in the implementation of periodic correlated methods.
%The computational scaling of these methods with the number of virtual states grows 
%between quadratically and exponentially, and therefore the ability to span the relevant parts of this space with as few functions as 
%possible becomes a critical design decision in the implementation of periodic correlated methods.

From the perspective of the paradigmatic uniform electron gas, plane waves are the natural choice of basis to expand both 
one-electron and many-electron wavefunction quantities~\cite{Shepherd2012,Shepherd2012a,Gruneis2013a,Shepherd2013a}.
These plane waves are 
% a natural orbital basis for both the mean-field and exact Hamiltonians, as well as being 
% I suppose they can be eigenfunctions of the mean-field (if there is no symmetry breaking), but they might not be
% especially natural for the exact Hamiltonian.
eigenfunctions of the kinetic energy operator, and naturally fulfil the periodicity of the computational cell. In realistic {\em ab initio} calculations, 
these plane waves also have a number of appealing features. They are independent of the molecular makeup
of the simulation cell, and instead only depend on the size and geometry of the cell. In addition, a single cutoff parameter dictating 
the upper energy scale of the included plane waves (and hence the resolution of the resultant wavefunctions) is used to systematically 
expand the plane wave basis to completeness, in a fashion free from basis-set superposition error (BSSE)~\cite{Marsman2009,Gulans2014}.

However, there are some drawbacks to plane wave expansions. Since they make no reference to the nature of the atomic environment, they 
have equal basis coverage throughout the cell. This can lead to a great deal of wasted computational effort when studying defects, surfaces 
or lower-dimensional systems. This is due to the necessity for large amounts of vacuum in the cells to minimize the effect of spurious periodic 
images, which results in large, unwieldy plane wave expansions to converge relevant properties~\cite{Gruneis2011a}. Even in bulk systems, 
the dominant electron density will generally be clustered around the atomic sites, and making use of this fact with atom-centered functions 
can certainly improve basis set convergence. 

For correlated methods, the steeper scaling compared to mean-field methods means that the speed of convergence with respect to number of basis functions
is even more critical. Rather than using plane waves directly for the virtual space, it is more common to truncate the prior mean-field virtual manifold 
on an energetic criteria to improve the convergence to the complete basis set\cite{Booth2013,PhysRevLett.114.226401}. 
However, this virtual manifold is not inherently physical, and its somewhat 
arbitrary truncation therefore does not necessarily provide a good basis in which to expand the correlated wavefunction. Truncating based on an 
energetic criteria (now on the mean-field energy rather than kinetic energy as for plane waves) does not necessarily yield fast convergence, while issues 
such as band-crossings which can occur as unit cells are distorted or enlarged can yield discontinuities in potential energy surfaces and equations of state. 
Furthermore, consistent truncations of the virtual bandstructure when comparing fundamentally different systems, such as a defective and pristine lattice structure,
are close to impossible to achieve, and therefore still generally rely on convergence to costly, near-complete basis sets for meaningful comparisons.
Attempts to truncate virtual single-particle orbital expansions using other criteria, such as occupation numbers from other levels of theory, have had some success, but can often be expensive to carry out~\cite{Gruneis2011a}.

An alternative representation to plane wave expansions are local atom-centered basis sets. This mirrors the duality between basis representations of lattice 
models such as the Hubbard model, where the local `site' degrees of freedom contrast with that of the discrete $k$-space, plane wave representation, 
commonly used in the uniform electron gas. These local functions now correspond to a basis with a local, atomistic description of the simulation cell.
Futhermore, their use allows for simple extraction of local descriptors, such as atomic electron numbers, spin density or projected densities of states, without requiring post-processing localization
steps towards Wannier, or similar, functions~\cite{Schmidt15,Louie15}.
While these local functions can take many forms, from Muffin-tin orbitals~\cite{Andersen75,Kotani2010,Exciting}, to wavelets~\cite{BigDFT,Madness,RevModPhys.71.267}
or numerical atomic orbitals~\cite{FHIAims,Siesta}, in this manuscript we consider the use of 
periodic Gaussian basis sets, and analyse their convergence for correlated levels of theory. 

While Gaussian basis sets can potentially take many 
different parameterizations, their widespread use within the field of quantum chemistry has meant that many tabulated basis sets
of increasing size and flexibility are readily available~\cite{Dunning89,Peterson05}. 
The Gaussian orbitals are optimized to approximate the natural orbitals of the free atom and its common ions, 
often at correlated levels of theory. Orbitals beyond the core and valence shells are included
to account for appropriate polarization and distortion of the atomic wavefunctions in bonding environments,
and to provide a description of correlation effects. 
Basis sets are commonly arranged in hierarchies so that they can 
be systematically expanded to allow for consistent and (if necessary) extrapolatable convergence.
In periodic systems, as the atomic-like Gaussian orbitals come together to form bands, they will
split about the Fermi level to describe the important low energy regions of the space, as well as retaining a consistent, local description of the one-electron 
wavefunctions, even for low-dimensional systems, or as cells change shape or atoms move.  The use of Gaussian-type orbitals in periodic electronic structure is not new to this work, with several other codes employing their use~\cite{Crystal,Cryscor,LocalMP2,Turbomole,Parinello99,Parrinello2000,Scuseria2000,Pulay2002,Vandevondele03}. The local nature of these functions and `nearsightedness' of the interactions 
is often used for reduced scaling techniques, including in diagonalization steps, or construction of Coulomb and exchange interactions in order to approach linear 
scaling mean-field treatments~\cite{LinearScalingMethods,Frisch96,Frisch96_2,Scuseria03}, and can also be extended to local treatment of correlation~\cite{LocalMP2,LocalMP2_2}. Furthermore, mixed plane-wave and Gaussian schemes have also been previously introduced as an attempt to combine their strengths in the condensed phase \cite{DelBen2012}. 

It should be noted that in post-mean-field correlated methods (including those based around the explicitly screened Coulomb interaction) the ultimate rate determining 
scaling in the convergence of correlated properties with respect to the one-electron basis set size is the description of the short-range Coulomb hole and non-analytic 
cusp condition at the coalescence point of two particles~\cite{Kato:CPAM10-151,Kutzelnigg}. It has been shown that this scaling behaviour is the same for both 
Gaussian, as well as plane wave expansions of the orbital space~\cite{Kutzelnigg,Shepherd2012}. However, the absolute convergence in different basis sets can be very 
different, as the decay of the Coulomb hole depends sensitively on the electron density, as well as the flexibility of the basis at the coalescence points. Furthermore, 
the absolute one-electron basis set incompleteness (which is both a feature of the correlated, as well as mean-field wavefunctions), is also very much dependent on 
the specifics of the primitive orbital expansion~\cite{Booth2012_2}.

In this paper, we detail an implementation of a straightforward approach for the use of a periodic Gaussian basis
(or indeed any numerical atom-centered functions) within a code set up for more traditional 
plane wave description of the wavefunction. We also consider changes to deal with core electrons when they are not explicitly considered, as is the case in the {\tt VASP} code within the Projector Augmented 
Wave (PAW) framework where this work is implemented. We then apply correlated levels of theory to a number of systems, demonstrating that the consistent level of truncation as afforded by the Gaussian basis 
set expansions leads to a rapidly convergent and extrapolatable virtual space for the calculations. Extensive comparison is made to all-electron molecular calculations, giving confidence in the applicability of the functions for
both strongly and weakly interacting systems.

%This is compared to the corresponding energetic truncation of plane wave or mean-field orbitals. 
%This reduction in virtual space is shown to be especially true when considering defects or surface physics, where no vacuum is explicitly considered in the virtual basis. 
%In addition, cohesive energies are obtained, where the consistent level of accuracy between the bulk and atomic descriptions lead to a rapidly convergent expansion.
For larger-scale applications, we propose and explore an efficient hybrid approach. In this approach, the occupied orbitals are converged first within a large primitive 
plane wave expansion, rendering the occupied space and hence Hartree--Fock energy and its contributions to properties essentially complete. A virtual
basis is then included for the correlation treatment comprised of the complementary set of orbitals constructed by projecting
the occupied orbitals out of the Gaussian basis set.
This dual approach removes basis incompleteness of the occupied one-electron wavefunctions and properties and
thus substantially ameliorates the issue of basis-set superposition error, which now only manifests through 
the subsequent correlation treatment. In addition, it retains the benefits of
the compact, consistent virtual space afforded by the atom-centered Gaussians.
% and removing the need for an energetic basis truncation. 
We note that related occupied projected Gaussian bases (projected atomic orbitals~\cite{pulay}) have been
used previously in Gaussian basis codes to exploit the locality of correlation\cite{Cryscor,LocalMP2_2}, however we do not
rely on this locality here, beyond its manifestation in the general compactness of the 
full set of virtual orbitals, as discussed above.
This is applied to the MP2 contributions to the cohesive energy of the Neon solid, where the weak binding means that basis set incompleteness 
manifests as large relative errors. Also studied is the absorption of a water molecule onto a Lithium Hydride crystal surface, where the low-dimensionality of
the system means that the Gaussian virtual space efficiently spans the correlated wavefunction, and allows
for rapid convergence of extrapolations which agree well with reference results.

\section{Construction of Gaussian basis}

%-------------------------------------------------------------------------
\begin{figure}[t]
    \begin{center}
        \includegraphics[width=8.0cm,clip=true]{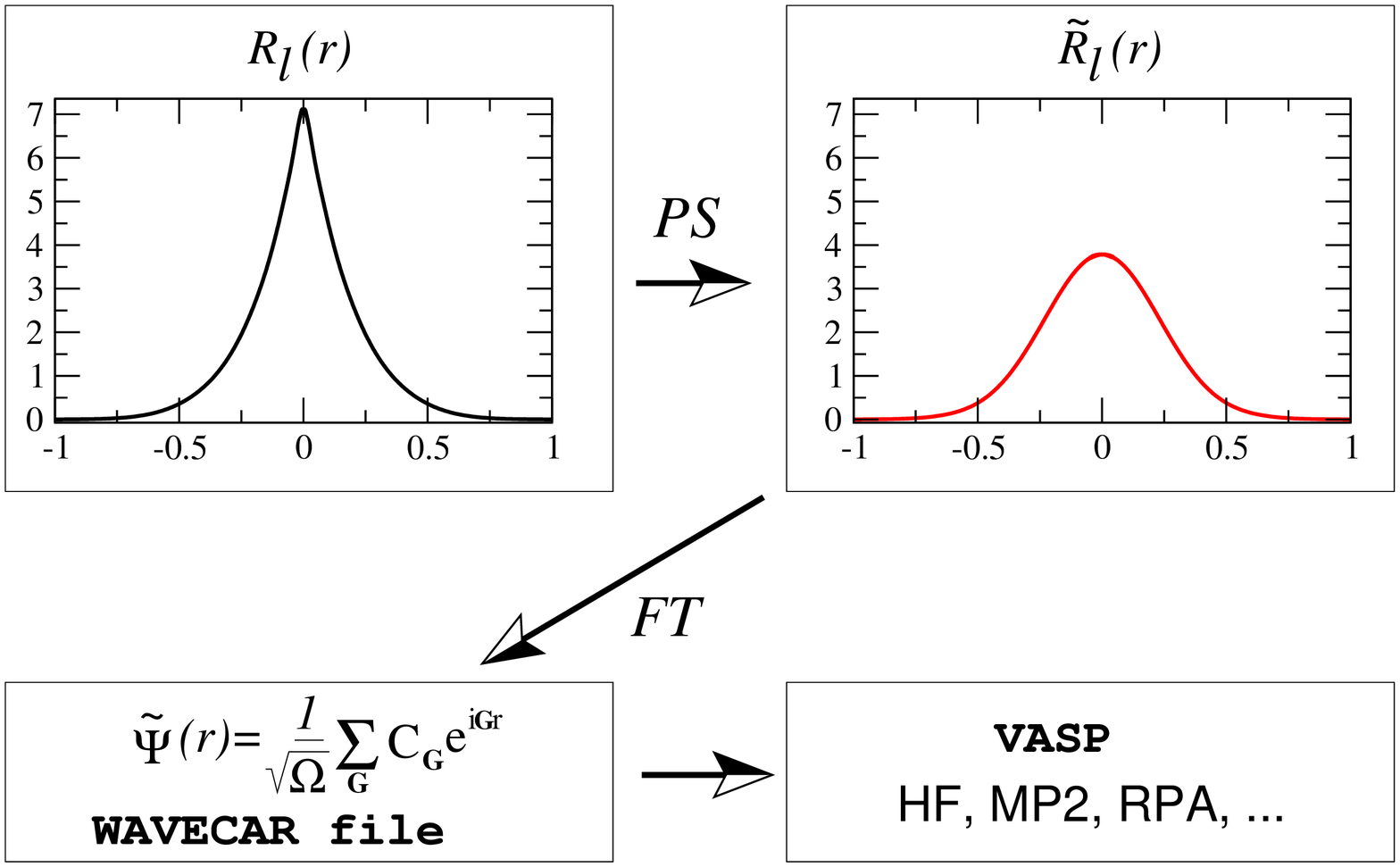}
    \end{center}
    \caption{Schematic illustration of the workflow. A Gaussian basis set is represented on a
radial grid (left upper panel). We employ a pseudization strategy (PS) to allow the core to be more efficiently represented by its Fourier coefficients by 
calculating the corresponding pseudized basis function (right upper panel). In the following step we Fourier transform (FT) the
{\it pseudized} Gausssian-type orbital (PGTO) to the plane wave basis (left bottom panel). This is then used for the various
electronic structure methods ranging from Hartree--Fock (HF) to post-HF theories.
%Finally we employ the PGTO orbital basis represented in a PW basis (stored in the WAVECAR
%file of \texttt{VASP}) for the various electronic structure methods ranging from HF to post-HF theories.
    }
    \label{fig:Scheme}
\end{figure}
%-------------------------------------------------------------------------

This section outlines the procedures employed to transform a
specified contracted Gaussian basis set into one which can be used with a plane wave solid state code within
the framework of the projector augmented wave method. This involves first `pseudizing' the sharply peaked core 
part of the basis,
designed to capture the nuclear cusp within all-electron calculations, but which is unnecessary in this context.
The resulting smoother function is then represented through its plane wave coefficients. 
Figure~\ref{fig:Scheme} outlines the individual steps schematically.

\subsubsection{Gaussian basis functions}

%I took some of the following equations simply from wikipedia. 
The Gaussian basis set is composed of atom-centered functions (GTOs) that can be decomposed into
a radial and angular parts such that an angular momentum function ($m,l$) for a given atom positioned at $\bf R$ 
can be given by 
\begin{equation}
G_{m,l,\bf R}({\bf r})=R_{l}(|{\bf r-R}|)Y_{l,m}(\Theta,\phi),
\end{equation}
where $Y_{lm}(\Theta,\phi)$ is a spherical harmonic and ${\bf r},\Theta,\phi$ correspond to spherical coordinates.
The radial function $R_l(|{\bf r-R}|)$ is expanded using Gaussian functions such that
\begin{equation}
R_{l}(|{\bf r-R}|)=|{\bf r-R}|^l\sum_{p}c_p A(l,\alpha_p)e^{-\alpha_p |{\bf r-R}|^2},
\label{eq:radgto}
\end{equation}
where $A(l,\alpha_p)$ is a normalization constant of the Gaussian primitives.
$c_p$ are contraction coefficients for the primitive Gaussian functions with exponent $\alpha_p$.

\subsubsection{The PAW method\label{sec:paw}}

The projector augmented wave (PAW) method was introduced by Bl\"ochl~\cite{Blochl1994}.
Further details, as well as its close relationship with the ultrasoft pseudopotential method of Vanderbilt, 
were shown by Kresse and Joubert in Ref.~\onlinecite{Kresse1999}, while here we briefly recap the approach.
%Its close relationship to the ultrasoft pseudopotentials method of Vanderbilt
%was shown by Kresse and Joubert~\cite{Kresse1999}, where further details can be found, while here we briefly recap the approach.
In the PAW method, the orbitals ($| \psi_n \rangle$)
are derived from the pseudo orbitals ($| \tilde{\psi}_n \rangle$)
by means of a linear transformation
\begin{equation}
| \psi_n \rangle = | \tilde{\psi}_n \rangle + 
                   \sum_i ( | \varphi_i \rangle - | \tilde{\varphi}_i \rangle ) 
                   \left \langle \tilde{p}_i | \tilde{\psi}_n \right \rangle
\label{eq:paw}
\end{equation}

The index $n$, labeling the orbitals $\psi$,
is understood to be shorthand for the band index and the Bloch wave
vector $k_n$, while the index $i$ is a shorthand for the atomic site $R_i$,
the angular momentum quantum numbers $l_i$ and $m_i$, and an additional index $\epsilon_i$
denoting the linearization energy. The wave vector is conventionally chosen to lie
within the first Brillouin zone.
The pseudo orbitals are the variational quantities of the PAW method
and are expanded in reciprocal space using plane waves,
\begin{equation}
\langle {\bf r} | \tilde{\Psi}_n \rangle = 
      \frac{1}{\sqrt{\Omega}} \sum_{{\bf G}} C^n_{{\bf G}}  e^{i(k_n+{\bf G}) {\bf r}}
\label{eq:pw}
\end{equation}

The all-electron partial waves $\varphi_i$ are the solution
to the radial Schr\"odinger equation for the non-spin-polarized reference atom at
specific energies $\epsilon_i$ and specific angular momenta $l_i$.
The pseudo-partial waves $\tilde{\varphi}_i$, are equivalent to the all-electron partial waves
outside a core radius $r_c$ and match continuously onto $\varphi_i$ inside the core radius.
The partial waves $\varphi_i$ and $\tilde{\varphi}_i$
are represented on radial logarithmic grids, multiplied with spherical harmonics.
The projector functions $\tilde{p}_i$ are
constructed in such a way that they are dual to the pseudo partial waves, i.e.,
$\langle \tilde{p}_i|\tilde{\varphi}_j\rangle=\delta_{ij}$.
The pseudized partial waves $\tilde{\varphi}_i$ are obtained by pseudizing the 
all-electron partial waves $\varphi_i$ for a given  core radius $r_c$~\cite{Kresse1994}.

%Add detailed discussion about how the pseudization procedure to get $\tilde{\varphi}$ works.

\subsubsection{Pseudized Gaussians}

In this work we seek to employ Gaussian basis sets using a plane wave code.
Fourier components of strongly localized real space orbitals
decay very slowly, which manifests as a slow convergence of the orbitals with respect to the underlying plane wave energy cutoff. 
This cutoff energy dictates the size of the plane wave basis employed in Eq.~\ref{eq:pw}.
The slow convergence of the orbitals with respect to this cutoff is mainly due to sharp features of the Gaussian-type orbitals (GTOs)
resulting from the fitting of the non-analytic cusp behaviour at the nuclear coalescence point, which
in non-relativistic quantum theory exhibits a derivative discontinuity in the wavefunction. However,
the plane-wave basis we employ is augmented within the projector augmented wave framework, which
includes a description of the atomic core region of each atom. This augmentation largely resolves these
sharply varying parts of the wavefunction.

To this end we `pseudize' the GTO functions defined in Eq.~\ref{eq:radgto},
which smooths the core region of each function, defined up to a pseudization radius from the nucleus, $r_c$. 
This is done in a way consistent with the symmetry and norm of the orbitals, and 
results in a more rapidly convergent set of Fourier components representing pseudized GTOs (PGTOs). 
The employed pseudization strategy mirrors the work of Kresse {\em et al} in the construction of pseudized partial waves~\cite{Kresse1994} for pseudopotentials.
The core of the pseudized radial Gaussian basis functions are expanded in three spherical Bessel functions such that
\begin{equation}
\tilde{R}_l(r)=\sum_{i=1}^3\alpha_i r j_l (q_i r)
\end{equation}
with $q_i$ chosen such that the value of the function, as well as logarithmic derivatives match at the cutoff radius,
\begin{equation}
\frac{\partial}{\partial r}\left [ ln R_l(r) \right ] |_{r=r_c}=
        \frac{\partial}{\partial r}[ ln(r j_l(q_i r))] |_{r=r_c}.
\end{equation}
Moreover, we require norm conservation of the PGTO such that
\begin{equation}
\int_0^{r_c} \tilde{R}_l(r)^2 dr = \int_0^{r_c} {R}_l(r)^2 dr.
\end{equation}
%\textcolor{red}{The function value must be matched as well as the log. derivative (note Eq. 6, 7 constitute only 2 conditions}.
We note that for $r\ge r_c$ the following condition holds ${R}_l(r)=\tilde{R}_l(r)$.
We choose the pseudization
radius such that it is identical to the cutoff radius used by the projectors $\tilde{p}_i$ in the PAW method. In this manner we ensure
that that core region of the pseudized Gaussians is augmented with additional terms that capture the oscillatory and sharp features of the one-electron
wavefunctions in this region. Once the appropriate Fourier components are found, integrals between the orbitals can be obtained in the reciprocal basis
as normal, with the Gygi-Baldereschi scheme used to correct for the divergence at ${\bf G}=0$ of the Coulomb kernel in reciprocal space\cite{Gygi,Gruneis2013a}.
%Spherical Bessel functions span a natural basis for smooth pseudo functions.
%Once the PGTO are calculated from the GTOs as described above we proceed.

%-------------------------------------------------------------------------
\begin{figure}[t]
    \begin{center}
        \includegraphics[width=9.0cm,clip=true]{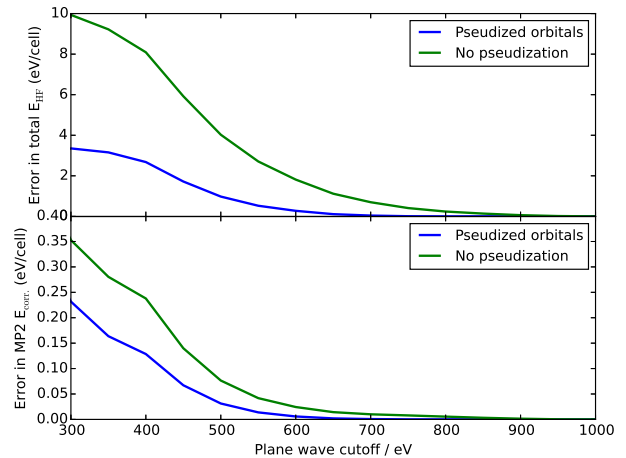}
    \end{center}
    \caption{
        The effect of pseudization on the convergence of Gaussian orbitals as plane waves, for a 3$\times$3$\times$3 simulation cell of diamond. The upper plot
        shows the difference between the Hartree--Fock total energy for the pseudized and non-pseudized STO-3G Gaussian orbital basis. This is calculated as
        the difference between a plane wave expansion of the orbitals truncated at a given energy and the `complete' (1000eV cutoff) description of the same orbital space. 
        Note that the 1000eV cutoff energies are not the same, since the act of pseudization slightly alters the orbitals.
        The lower plot shows the convergence of the MP2 correlation energy for the same system.
        The 1s orbitals are removed from the STO-3G atomic basis on each atom, since the 1s electrons are replaced by a pseudopotential, with the orbitals 
        represented within the PAW functions.
    }
    \label{fig:FourierCom}
\end{figure}
%-------------------------------------------------------------------------

\begin{figure}
    \begin{center}
        \includegraphics[width=9.0cm,clip=true]{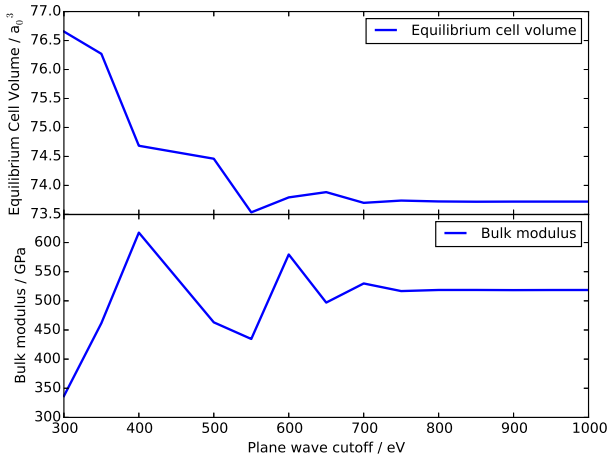}
    \end{center}
    \caption{Convergence of the equilibrium cell volume (upper plot) and bulk modulus (lower plot) with increasing plane wave cutoff, as obtained from the 
        equation of state of a $3\times3\times3$ cell of diamond. This equation of state was calculated at the MP2 level of theory in a cc-pVDZ pseudized Gaussian basis set, and
        fit to a Birch-Murnaghan form for 14 cell volumes at each plane wave truncation~\cite{PhysRev.71.809}.}
    \label{fig:EOSConv}
\end{figure}

A demonstration of the importance of the pseudization of the atomic Gaussian functions is given in Fig.~\ref{fig:FourierCom},
where the convergence of the total and correlation energies of a $3\times3\times3$ cell of diamond is considered, demonstrating that pseudization of
the atomic functions is essential to obtain rapidly convergent properties with the size of the underlying plane wave basis. While the size of this
plane wave basis is generally insignificant when considering the cost of the correlated treatment in the system, numerical and computational difficulties can arise if the
plane wave cutoff is too large, and therefore the pseudization is necessary when aiming to converge results for a given atomic basis set. We can also consider the
convergence of static properties in a larger Gaussian basis as the underlying plane wave basis which represents these functions increases, which will depend on the accuracy of relative energies across a range of cell geometries.
This is shown in Fig.~\ref{fig:EOSConv}, demonstrating that cutoffs of ~750eV are sufficient to saturate the
representation of the pseudized Gaussian orbitals, and to fully converge the equation of state for the system.

%-------------------------------------------------------------------------
%-------------------------------------------------------------------------

\subsubsection{Virtual pseudized Gaussian basis\label{sec:append}}

In instances where the local nature of the basis as Gaussians is not specifically required
in the mean-field calculation, it
can be highly beneficial to pursue a hybrid approach to the basis construction for correlated calculations.
In this, the relatively cheap mean-field part of the calculation can be performed with a plane wave basis, where 
convergence to the complete basis (e.g. to ~1meV in energies) can be achieved using affordable energy cutoffs,
and without the basis set linear-dependence that can be a problem in larger Gaussian basis sets.
% I have removed the exponentially fast part, since it is exponentially fast also in a Gaussian basis.
% The key thing is that with today's computers and desired accuracies it is just practical to carry out calculations
% at the basis set limit with PW's.
Once converged, this yields close to the optimal occupied orbitals, mean-field (Hartree--Fock) energy and wavefunction, 
essentially free from basis set incompleteness or superposition errors. This is advantageous because the Gaussian basis will not
in general span the full occupied orbital space, due to the polarization of the orbitals from to the environment.

The large expansion of virtual orbitals as plane waves or virtual canonical orbitals
is then avoided by representing the virtual space as an expansion of pseudized Gaussian basis functions,
after having projected out the component of the (complete) occupied space from the Gaussian basis. This ensures that the
full flexibility of the pseudized Gaussian basis is spanned, {\em in addition} to the complete occupied space. Since all
orbitals are ultimately expressed by their Fourier components, this projection is trivially achieved, and the virtual 
orbital space $\{ |\psi_{\alpha} \rangle \}$ can be constructed as
\begin{equation}
    | \psi_{\alpha} \rangle = | G_{\alpha} \rangle - \sum_i | \psi_i \rangle \langle \psi_i | G_{\alpha} \rangle
\end{equation}
where $| G_{\alpha} \rangle$ is the Gaussian basis, and $ | \psi_i \rangle $ represents the complete space of occupied orbitals
expressed in the plane wave basis (note that the $\psi_i$ orbitals which are projected out refer to the `all-electron' orbitals 
rather than the pseudized orbitals to ensure true orthogonality).
If the norm of any virtual orbital is below a threshold value after this projection, then it is removed from the calculation, while
the rest are orthonormalized and constitute the virtual basis for the calculation. The virtual space is subsequently canonicalized before use in 
post-mean-field methods, with no further mixing between occupied and virtual states possible.
It is also possible for the `occupied' atomic orbitals of the original contracted
Gaussian basis to be identified and removed entirely from the basis, as they are largely redundant, to leave an overall basis 
the same size as the original, unmodified Gaussian basis, but still complete in all mean-field orbitals and properties. 

The benefits of this basis construction are significant, with basis set incompleteness and superposition error only remaining in the
correlated treatment of the wavefunction, which is readily extrapolatable within the employed correlation-consistent basis sets.
%%  Furthermore linear dependencies
%% between more diffuse Gaussian basis functions, which typically arise in densely packed solid can be avoided in this approach at least on the level of
%% the mean-field energy.
%% This is now moved up. 
A drawback of the above construction is that the occupied space is no longer represented within an underlying local basis, which
may be desirable for the exploitation of locality approximations in quantum cluster methods. Finding a local representation would then
require further localization steps which would be unnecessary if the underlying basis was already local. The use of Gaussian basis sets
for local, cluster approximations will be explored in the future. We now turn to some applications to demonstrate the performance of the
pseudized Gaussian basis compared to all-electron molecular calculations, before a study of more challenging systems.

\section{Results}

\subsection{Comparison to molecular systems}

\subsubsection{He in a pseudized aug-cc-pVTZ basis set}

%-------------------------------------------------------------------------
\begin{figure}[t]
    \begin{center}
        \includegraphics[width=8.0cm,clip=true]{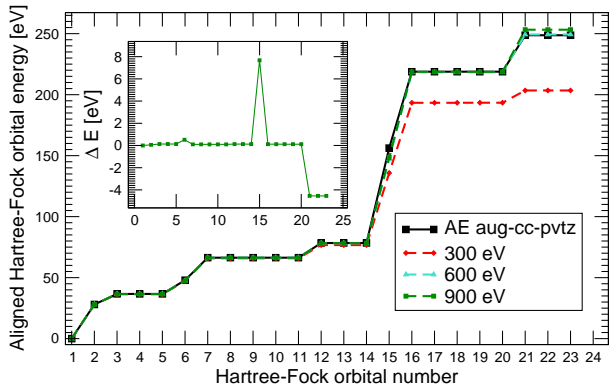}
    \end{center}
    \caption{Hartree--Fock (HF) one-electron energies retrieved as a function
of the orbital number for the He atom. The all-electron (AE) HF orbital energies using the
 aug-cc-pVTZ  basis set have been obtained using the \texttt{PSI4} quantum chemistry package.
The HF plane wave calculations using \texttt{VASP} have been performed in the same basis using a
20~$\times$20~$\times$20~\AA$^3$~cubic box.
    }
    \label{fig:1}
\end{figure}
%-------------------------------------------------------------------------

In order to quantify the effect of the pseudization, and to assess the fidelity of the representation of the Gaussian type orbitals, we first
compare to gas phase molecular systems, where results within the same basis 
obtained from a molecular Gaussian basis code can be compared to our periodic implementation in the
limit of a large simulation box. Here we use \texttt{PSI4}~\cite{Psi4} as the molecular code while all
pseudized GTO calculations were performed using \texttt{VASP}. The first investigation assesses the Hartree--Fock
one-electron energies of a He atom in a 20$\times$20$\times$20~\AA$^3$~cubic box.
The He atom in the aug-cc-pvtz basis set constitutes a test case because it is free
of frozen core states that might introduce an additional source of discrepancy between the \texttt{VASP} pseudopotential
and \texttt{PSI4} all-electron quantum chemistry results.
For the sake of clarity we will refer to the results obtained using the \texttt{VASP} code and pseudized GTOs as PGTOs results,
whereas the results obtained using \texttt{PSI4} will be referred to as GTOs results.
%Details of the employed He PAW potential are summarized in table~\ref{tab:HePAW}.
Figure~\ref{fig:1} shows the Hartree--Fock (HF) one-electron energies calculated using 300~eV, 600~eV and 900~eV plane
wave cutoff energies.
It can be seen that the HF one-electron energies converge rapidly for both occupied and virtual manifolds, 
even for the more high-lying states. It should be noted that both the virtual and occupied space was expressed in the underlying Gaussian basis in this example,
rather than using the technique detailed in section~\ref{sec:append}.

%The black line shows the HF one-electron energies calculated using GTOs.
The order and degeneracy of the states agrees between the two different methods (PGTOs and GTOs).
However, the inset in Figure~\ref{fig:1} shows that the energy differences
between the PGTOs and GTOs contain outliers corresponding to states 15 and 21-23, where the discrepancy can become larger than 4~eV. We attribute this 
to the different form of the PGTOs and the GTOs inside the PAW sphere.
In this region the PGTOs are pseudized and augmented with terms that depend on the projectors, the all-electron partial waves and the pseudo orbitals.
However, we stress that the aim of this work is not to achieve perfect agreement between
PGTOs and GTOs but rather to obtain a similar quality and basis set convergence for correlated
wave function calculations.
%\textcolor{red}{Could you check this hypothesis by changing the pseudization radius? Otherwise it doesn't seem altogether convincing.}
%It is clear that the convergence of correlation energies is very sensitive to
%the HF HOMO-LUMO

\subsubsection{Water dimer}
%-------------------------------------------------------------------------
\begin{figure}[t]
    \begin{center}
        \includegraphics[width=3.0cm,clip=true]{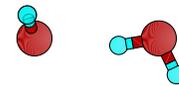}
    \end{center}
    \caption{Water dimer geometry.
    }
    \label{fig:water}
\end{figure}
%-------------------------------------------------------------------------

%-------------------------------------------------------------------------
\begin{table}[t]
\caption{
Atomic positions for the water dimer\footnote{http://cccbdb.nist.gov/} (\AA).
}
\label{tab:waterstruct}
\begin{ruledtabular}
\begin{tabular}{lccc}
       &   x  &  y  &  z  \\ \hline
O  &    0.004 &  1.491 &  0.000 \\
H  &    0.168 &  0.559 &  0.000 \\
H &     0.850 &  1.899 &  0.000 \\
O  &    0.004 &  -1.405 & 0.000 \\
H  &    -0.542 & -1.573 & 0.749 \\
H  &    -0.542 & -1.573 & -0.749 \\
%O & -0.002 &	1.517 &	0.000 \\
%O & -0.002 &	-1.386 &	0.000 \\
%H & 0.090 &	0.555 &	0.000 \\ 
%H & 0.898 &	1.846 &	0.000 \\
%H & -0.482 &	-1.723 &	0.759 \\
%H & -0.482 &	-1.723 &	-0.759 \\
\end{tabular}
\end{ruledtabular}
\end{table}
%-------------------------------------------------------------------------

%-------------------------------------------------------------------------
\begin{table}[t]
\caption{
Binding energy contributions for the water dimer using HF and MP2 theory, comparing results from gas-phase calculations in a contracted
aug-cc-pVXZ basis with frozen Oxygen 1s core, and
our results from a periodic system in a 10$\times$10$\times$10~\AA$^3$~cubic box, with a pseudized basis. The $(CP)$ denotes that
counterpoise corrections have been included for basis set superposition error in this basis.
%The first column defines the cardinal number of the aug-cc-pVXZ basis set employed.
%CP  in the first column stands for counterpoise correction and pecifies if the obtained binding energies haven been
%corrected for basis set superposition errors.
All units are in meV.
}
\label{tab:waterbinding}
\begin{ruledtabular}
\begin{tabular}{lccc|ccc}
%& \multicolumn{3}{c}{\texttt{VASP}} &\multicolumn{3}{c}{\texttt{PSI4}} \\
& \multicolumn{3}{c}{PGTOs} &\multicolumn{3}{c}{GTOs} \\
 Basis & HF  & MP2 corr.  & MP2 &  HF  & MP2 corr. & MP2  \\ \hline
 aVDZ & 142 & 56    & 197   & 154.9  & 62.4        & 217.3 \\
aVTZ & 142 & 66    & 208   & 145.4  & 68.6        & 214.9 \\
aVDZ (CP) & 142 & 41      & 183 & 151.2  & 42.5        & 193.7  \\
aVTZ (CP) & 142 & 53      & 195 & 143.8  & 52.8        & 196.6  \\
%aVDZ & 142 & 56    & 197   & 156  & 63        & 219 \\
%aVTZ & 142 & 66    & 208   & 146  & 88        & 234 \\
%aVDZ (CP) & 142 & 41      & 183 & 151  & 44        & 195  \\
%aVTZ (CP) & 142 & 53      & 195 & 144  & 51        & 195  \\
\end{tabular}
\end{ruledtabular}
\end{table}
%-------------------------------------------------------------------------

%-------------------------------------------------------------------------
%\begin{table}[t]
%\caption{
%Binding energy contributions for the water dimer using HF and MP2 theory including counterpoise corrections
%for the basis set superposition error. The
%first column defines the cardinal number of the aug-cc-pVXZ basis set employed.
%All units in meV.
%}
%\label{tab:waterbindingcp}
%\begin{ruledtabular}
%\begin{tabular}{lcccccc}
%& \multicolumn{3}{c}{{PGTOs}} &\multicolumn{3}{c}{{GTOs}} \\
% Basis & HF  & MP2 corr.  & MP2 &  HF  & MP2 corr. & MP2  \\ \hline

%Q & 142 &            &     & 145  & 62        & 207   \\
%\end{tabular}
%\end{ruledtabular}
%\end{table}
%-------------------------------------------------------------------------

Having demonstrated that the PGTOs yield a similar one-electron spectrum as the GTOs, as well as rapidly convergent
properties with plane wave cutoff, we now turn to
the discussion of the convergence of correlation energy contributions to the binding energy in the water
dimer with basis size. Figure~\ref{fig:water} and Table~\ref{tab:waterstruct} show and specify the employed water dimer
structure. 
This case is important because plane wave basis set calculations are computationally
very expensive for atomic and molecular systems, which usually require a large simulation
box with a lot of vacuum to minimize interactions between periodic images, and are required for many formation or cohesive energy calculations. 
As a result of the large
box size the number of plane wave basis functions becomes very large regardless of the actual number
of electrons in the unit cell~\cite{Preuss04}.
In this calculation we have employed a 10$\times$10$\times$10~\AA$^3$~cubic box.
The cutoff energy was set to 600~eV resulting in a basis set size consisting of more than 30,000 plane
waves, which would be impossible to do at the level of MP2. However, the number of Gaussian basis functions is smaller than 200 even for the pseudized
aug-cc-pVTZ basis set.

For this example (and subsequent applications in this work), the complete occupied space of states was first calculated, and included within 
the basis to saturate the Hartree--Fock wavefunction. The PGTO basis was therefore only
used to span the complementary virtual space, as described in Sec.~\ref{sec:append}.
We note that the MP2 results for this system change by less than 3~meV when employing a
15~$\times$15~$\times$15~\AA$^3$~cubic box. This is unsurprising, since even the most diffuse Oxygen functions in the 
aug-cc-pVTZ basis have less than 0.001\% of their integrated radial density found more than 6\AA~ away from their nuclear origin.
This locality of basis coverage is the dominant factor in the compactness of the Gaussian basis for this purpose compared to
plane wave expansions. We note here that in all calculations, a standard contracted Gaussian basis was used for the construction of the PGTOs.
%VASP calculations using a 10x10x10 box. MP2 results change by less than 3 meV using a 15x15x15 box.
%LAPPEND was used.

Table~\ref{tab:waterbinding} summarizes the obtained binding energies for the water dimer
on the level of HF and MP2 theory calculated using the (pseudized) aug-cc-pVDZ and aug-cc-pVTZ contracted basis sets. The Oxygen 1s orbitals of the
GTO calculations were frozen to allow for fairer comparison to the valence-only PGTO calculations performed with the PAW method.
Since the PGTOs are only used to span the virtual orbitals, the PGTOs
Hartree--Fock contribution to the binding energies are independent of the basis set size, since these are necessarily complete and free of basis set errors.
This also renders the occupied space and HF contribution free of basis set superposition error, as shown by the lack of a counterpoise correction to the values.
It can be seen that the HF contribution from the true, gas-phase GTOs approaches that of the PGTOs as the basis is increased.

%In the case of the aug-cc-pVTZ basis set GTOs and PGTOs yield HF binding energies of 142~meV and 146~meV, respectively.
%Tables~\ref{tab:waterbinding} and~\ref{tab:waterbindingcp}
%summarize results that have been obtained without and with counterpoise corrections for the basis set superposition error.
Table~\ref{tab:waterbinding} shows that the PGTOs and GTOs yield MP2 binding energies that deviate by more than
20~meV from each other when no allowance is made for basis set superposition error. However, the majority of this difference is due to the
incompleteness in the HF contribution to the interaction in the GTO basis. The discrepancy in the MP2 correlation contribution to the interaction energy is
only at most 6~meV, despite the contrasting occupied space. This difference decreases as the basis set is increased and the HF contribution of the GTO basis
becomes increasingly complete to match the PGTO calculations.
%This corresponds to approximately 
%10\% of the total binding energy even though the same basis set type is used.
%We note that these results have been obtained without accounting for the basis set superposition error.
Basis set superposition errors are a well-known drawback of atom-centered basis sets and can typically lead to
an overestimation of binding energies. This is because more basis coverage is available for each monomer at shorter bond lengths, as it can exploit the
basis coverage supplied by the overlap of the functions from the other monomer. 

To better understand the origin of the difference between PGTOs and GTOs
we have also included in table~\ref{tab:waterbinding} calculations with counterpoise corrections for the basis set superposition error (BSSE).
%These results are summarized in Table~\ref{tab:waterbindingcp}.
Both our PGTO and GTO calculations lower the predicted binding
energies by about 15~meV and 20~meV in the case of PGTOs and GTOs respectively if BSSE is accounted for.
This basis set superposition error in the case of the PGTO is purely contained in the correlation part of the MP2 since the occupied space is complete, while for the GTO basis
it also includes basis set superposition in the HF contribution. Once the BSSE of purely the correlation part of the MP2 interaction energy
is analysed, the errors are in closer agreement, however, these will still be affected by the contrasting occupied space in each system.
%Furthermore the discrepancy between PGTOs and GTOs for the MP2 correlation energy contribution to the binding energy is significantly
%reduced and is on the order of 2-3~meV in the case of the aVDZ and aVTZ basis set.
From this example we conclude that our PGTO basis sets yield a comparably accurate description of electronic correlation effects as
the GTO counterparts for molecular systems. 
%However, the observed BSSEs in PGTOs and GTOs are different, indicating that
%the procedure employed in constructing virtual PGTOs as outlined in Sec.~\ref{sec:append} results in slightly smaller BSSEs in PGTOs as
%compared to GTOs for this system.
%the agreement between the best basis sets available is equally good.
%\texttt{PSI4} and \texttt{VASP} yield MP2 binding energies of 207~meV and 210~meV, respectively.
%At the moment we have not implemented $g$-functions in the PGTOs module of \texttt{VASP}, which restricts
%our basis set calculations to triple Zeta quality. However, 
We note that the achieved accuracy of a few meV provides confidence in the constructed basis, and is sufficiently accurate 
to allow for reliable predictions in $ab~initio$ calculations.

%% UP TO HERE %%

\subsubsection{Nitrogen molecule}

%-------------------------------------------------------------------------
\begin{table}[t]
\caption{
Binding energy contributions for the Nitrogen molecule using HF and MP2 theory, for all-electron, gas-phase GTO results, and our contracted PGTO implementation.
The first column defines the type of contracted aug-cc-pVXZ basis set employed.
$(CP)$ denotes that the results have been corrected for basis set superposition error via counterpoise correction.
All units are in eV.
}
\label{tab:nitrogenbinding}
\begin{ruledtabular}
\begin{tabular}{lccc|ccc}
& \multicolumn{3}{c}{{PGTOs}} &\multicolumn{3}{c}{GTOs} \\
Basis         &   HF  & MP2 corr. & MP2    &  HF    &  MP2 corr. & MP2  \\ \hline
aVDZ      & 5.094 &  4.153    & 9.247  & 4.754  & 4.335      & 9.089  \\
aVTZ      & 5.094 &  4.771    & 9.864  & 5.043  & 5.046      & 10.089  \\
aVQZ      & 5.094 &     -      &   -     & 5.096  & 5.111      & 10.207 \\
aVDZ (CP) & 5.094 &  3.944    & 9.038  & 4.722  & 4.157      & 8.879  \\
aVTZ (CP) & 5.094 &  4.674    & 9.768  & 5.033  & 4.731      & 9.764  \\
aVQZ (CP) & 5.094 &      -     &    -    & 5.094  & 5.004      & 10.097 \\    
%Q   & 5.079 &           &        & 5.094  & 5.004      & 10.097  \\
\end{tabular}
\end{ruledtabular}
\end{table}
%-------------------------------------------------------------------------

As a final comparison for molecular systems, we now seek to investigate the performance of the pseudized 
GTO basis for the atomization of a prototypical covalently bound system, the Nitrogen molecule.
The calculations of N$_2$ were performed using a 15$\times$15$\times$15~\AA$^{3}$~cubic unit cell
to minimize the interaction between the periodic images. The plane wave energy cut off was set to
600~eV. A bond length of 1.0656~\AA~was used as the equilibrium geometry.

Table~\ref{tab:nitrogenbinding} summarizes the obtained binding energies on the level of
HF and MP2 theory using PGTOs and GTOs, with and without counterpoise corrections for the BSSEs, with UHF/UMP2 used for the calculation of the atomic system in
each instance.
Again our results show that the PGTO Hartree--Fock contributions
to the dissociation energy are independent of basis size, and agree well with the GTO results from large basis sets. The GTO results using
aug-cc-pVQZ yield HF atomization energies that deviate by less than 2~meV from our
results obtained using PGTOs within \texttt{VASP}.
On the level of MP2 theory the PGTOs and GTOs results differ more strongly, especially for the aVDZ basis set.
This is a result of the number of differences: the PAW framework for the core electrons, the pseudization of the basis, and the different
virtual space, due to the construction detailed in section~\ref{sec:append} and orthogonalization to the complete occupied space.
However, as the basis is increased to aug-cc-pVTZ, the agreement is improved, likely to be due to the fact that the larger space mitigates the discrepancies
in the construction of the virtual space, as the occupied space of the GTOs is rapidly approaching completeness. Once corrections for BSSE
are also included, the agreement in the MP2 is very good, with the PGTOs and GTOs predicting a correlation energy contribution of 
4.674~eV and 4.731~eV to the binding energy, respectively.

%We attribute this discrepancy to the fact that even the HF dissociation energies are not very well converged
%for the aVDZ basis, as a result the orbitals and the corresponding correlation energies will differ as well between PGTOs and GTOs.
%In the case of the aVTZ basis set, the HF energies agree to within 2~meV and 61~meV without and with BSSE corrections, respectively.
%Furthermore the agreement of the MP2 correlation energy contribution to the binding energy between PGTOs and GTOs using the aVTZ and CP
%corrections is very good. PGTOs and GTOs predict a correlation energy contribution of  4.674~eV and 4.731~eV to the
%binding energy, respectively.
%We also note that the change in the MP2 correlation energy contribution to the binding energy from aVDZ to aVTZ is also very
%similar between PGTOs and GTOs if CP corrections are employed. PGTOs and GTOs predict an increase in the binding energy of
%730~meV and  
However, the correlation energy contribution to the binding energy converges very slowly with respect 
to the largest angular momentum quantum number included in the basis, and comparison to the aug-cc-VQZ GTO basis results shows that 
we are not yet converged to `chemical accuracy'.
Since we have not yet implemented the required transformation routines for $g$-functions, we cannot employ aug-cc-pVQZ PGTO basis sets or larger for this element. We also stress that explicitly correlated methods will greatly help
in yielding a more rapid convergence of the binding energies with respect to the basis set size for such covalently bonded systems.
These methods and their implementations are already being investigated in the framework of fully periodic systems using a plane
wave basis set and the combination of these different techniques will be subject to a future study~\cite{Gruneis2013a,PhysRevLett.115.066402}.
Nonetheless our results indicate that
the pseudized GTOs yield results that are very similar compared to the all-electron GTO results. While establishing the validity of
this comparison is important, the aim of this procedure is not for application to molecular systems, but rather for a compact basis
for extended systems, which we now consider and compare to experiment.

\subsection{Extended systems}

\subsubsection{Neon solid}

One area where a local Gaussian virtual space representation is expected to perform well compared to plane wave expansions
is in the description of weakly interacting, dispersion dominated extended systems. 
The contracted aug-cc-pV$X$Z hierarchy is expected to provide a rapidly convergent and systematic truncation of the virtual basis by
spanning a space constructed to obtain the required higher energy excitations for the dispersion interaction, as compared to a strict energetic
truncation of plane wave or canonical virtual orbitals.
%including elements of required higher energy excitations, as compared to strict energetic 
%truncations of plane wave or canonical virtual orbitals.
This systematic truncation can be of great benefit if one seeks to calculate converged energy differences between solids and isolated
atoms, where a strict orbital energetic truncation becomes physically meaningless.
To this end, we demonstrate the calculation of the atomization energy of the Neon noble gas solid.
The Neon solid has an fcc unit cell, with a lattice constant of 4.641\AA, and the pseudized Gaussian basis for the virtual space
is expanded in plane waves up to a cutoff parameter of 700eV. The MP2 results of the solid have been calculated using
a 6$\times$6$\times$6 $k$-point mesh,
%sampled cell are extrapolated
%to the thermodynamic limit using a $\frac{1}{N_k}$ extrapolation
while the box size for the atomic system is 
30$\times$30$\times$30~\AA$^3$.
For the HF contribution to the cohesive energy we choose an even denser $k$-point mesh of 14$\times$14$\times$14.
These parameters are sufficient to converge the cohesive energy to within 1~meV.

Results for the atomization energy of this system can be seen in Table~\ref{tab:neon}, where we present HF and MP2
results obtained using PGTOs with and without counterpoise (CP) corrections for the BSSE. 
Since the occupied space is the same for each basis, as the basis choice
simply affects the virtual orbitals in this scheme,
the Hartree--Fock contribution to the atomization energy in each basis is also seen to be the same. The
binding of the solid is also purely dispersive.
As dispersive interactions are a manifestation of correlated phenomena, this renders the Hartree--Fock contribution negative,
representing a repulsive interaction at this level of theory.
We therefore consider the atomization energy at the level of MP2 theory
(which is believed to describe dispersion interactions well in this system)
with different choices of virtual basis.

%Table~\ref{tab:neon} summarizes the HF and MP2 results.
The benchmark MP2 result for this system has been obtained using the incremental method, as detailed in Ref.~\onlinecite{Schwerdtfeger2010}, which relies
on a truncated many-body expansion for the interactions, and has been shown to work particularly well for such noble gas solids or molecular crystals~\cite{Incremental}.
%We summarize again our MP2 results obtained using PGTOs with and without counterpoise (CP) corrections for the BSSE.
Our findings indicate again that BSSEs can be quite large on a relative scale for these weakly bound systems and need to
be accounted for. If CP corrections are included, our aug-cc-pVDZ and aug-cc-pVTZ PGTO calculations predict MP2 atomization
energies of 10~meV and 16~meV, respectively. To correct for the remaining basis set incompleteness error we can also perform
a complete basis set (CBS) limit extrapolation, theoretically justified according to the $1/L^3$ convergence behavior of the correlation 
energy~\cite{Kutzelnigg92,Kutzelnigg92_2,Helgaker97}. The obtained CBS limit result is 19~meV, which is in good agreement with results obtained using the incremental method of 18.8~meV, 
although we believe that the remaining error from $k$-point sampling and convergence of other technical parameters in our calculations is
on the order of 1~meV, and therefore such
good agreement is somewhat fortuitous.
We note that the experimental value of the atomization energy corrected for zero-point vibrational effects is 27.8~meV~\cite{Schwerdtfeger2010,simmons67,simmons68,Endoh75},
demonstrating an underbinding of MP2, consistent with the trend observed for gas-phase noble gas dimers and of previous MP2 results for the Neon crystal\cite{Schwerdtfeger2010,Merkt03,Ogilvie92,Herman88}.
%\textcolor{red}{(on the order here sounds a bit imprecise, can we give the number with error bars?)},
%underlining the accuracy of MP2 theory for this system~\cite{Merkt03,Ogilvie92,Herman88}.

%-------------------------------------------------------------------------
\begin{table}[t]
\caption{
Cohesive energy of the Neon solid using HF and MP2 theory, in PGTO basis sets, with and without corrections for BSSE.
Comparison to the incremental results of Ref.~\onlinecite{Schwerdtfeger2010} for the MP2 energy are included.
All units are in meV.
}
\label{tab:neon}
\begin{ruledtabular}
\begin{tabular}{lcc}
 Basis      &    HF    & MP2      \\ \hline
aug-cc-pVDZ (no CP)   &   -8  & 19  \\
aug-cc-pVTZ (no CP)   &  -8   & 24   \\
aug-cc-pVDZ (CP)          &  -8   & 10  \\
aug-cc-pVTZ  (CP)         &  -8   & 16   \\
aug-cc-pV(D,T)Z  (extrap,CP)     &  -8   & 19     \\
Incremental method                &       &  18.8
\end{tabular}
\end{ruledtabular}
\end{table}
%-------------------------------------------------------------------------

\subsection{Water at LiH}

%-------------------------------------------------------------------------
\begin{figure}[t]
    \begin{center}
          \includegraphics[width=4.0cm,clip=true]{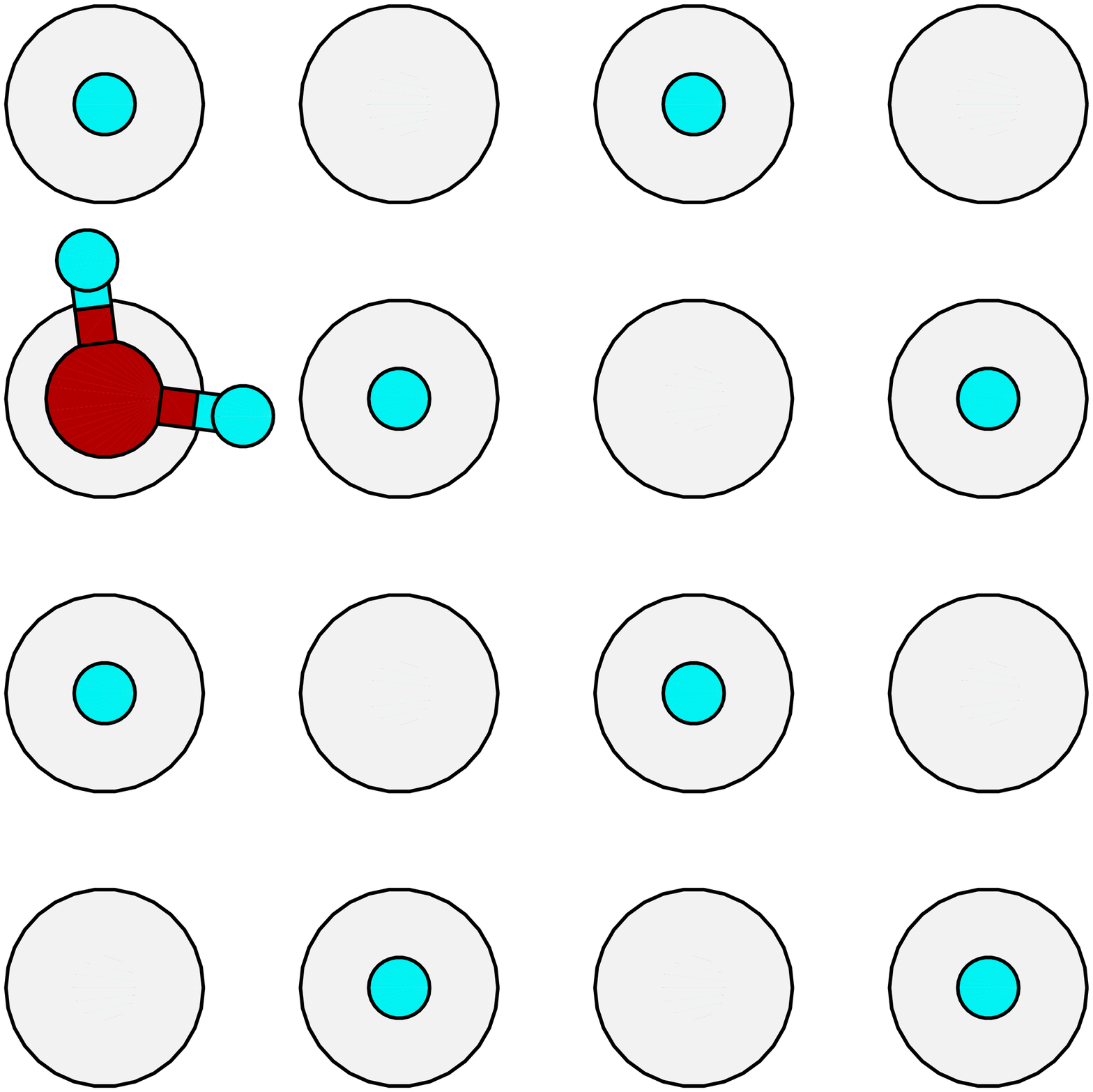}
          \includegraphics[width=4.0cm,clip=true]{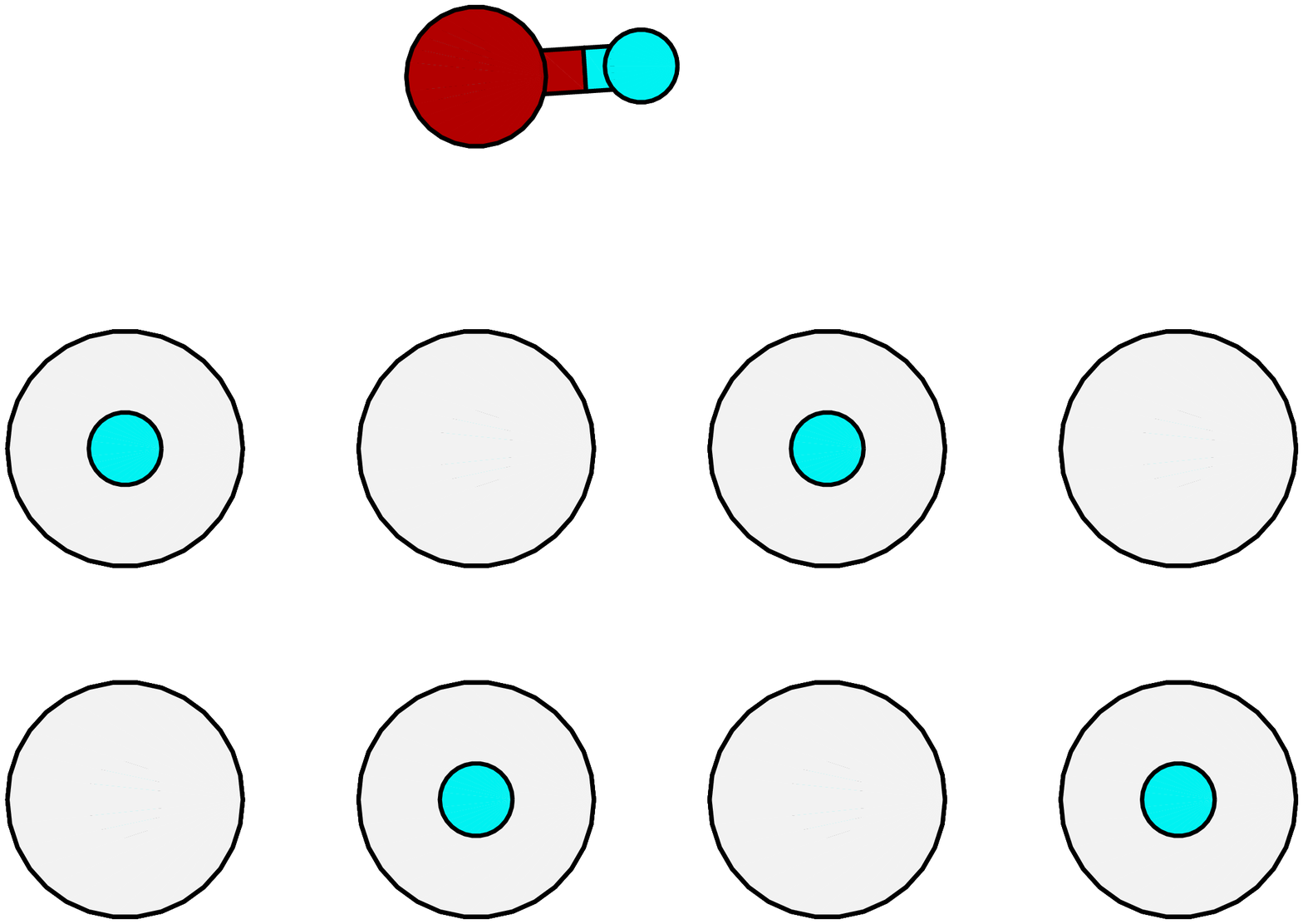}
    \end{center}
    \caption{H2O@LiH adsorbtion site studied. 
    }
    \label{fig:site}
\end{figure}
%-------------------------------------------------------------------------

%-------------------------------------------------------------------------
\begin{table}[t]
\caption{
Adsorption energies for one water molecule on the LiH surface at various levels of theory. Hartree--Fock and MP2 results
are performed in different PGTO basis sets, corrected for BSSE, and extrapolated to the CBS limit for comparison. Results from the incremental
method and DMC are taken from Ref.~\onlinecite{Binnie2011}.
%, RPA@PBE and RPA+SOSEX@PBE theory.
All units are in meV.
}
\label{tab:waterbinding2}
\begin{ruledtabular}
\begin{tabular}{lccc}
& \multicolumn{3}{c}{{$E_{\rm ads}$/meV}}   \\
              &  cc-pVDZ & cc-pVTZ  & CBS   \\ \hline
HF            &  44      & 44       & 44    \\
MP2           &  157     & 195      & 211   \\
%DCSD          & -165     &          &       \\
%Incremental [MP2]	&	&	&	215 \\
\hline
DFT-PBE       &     &     & 214 \\
Incremental [CCSD(T)]	&	&	&	246 \\
DMC	   		&	&	&	237 \\
%RPA           & 258 & 226   \\
%RPA+SOSEX     & 228 & 216  \\
\end{tabular}
\end{ruledtabular}
\end{table}
%-------------------------------------------------------------------------

%-------------------------------------------------------------------------
\begin{table}[t]
\caption{
Adsorption energies for one water molecule on the LiH surface using 
MP2 theory employing different virtual orbital manifolds with and without counterpoise corrections for the BSSE.
cc-pV(D,T)Z denotes the three-point extrapolation to the complete basis limit~\cite{Helgaker97}.
%, RPA@PBE and RPA+SOSEX@PBE theory.
All units are in meV.
}
\label{tab:waterbinding3}
\begin{ruledtabular}
\begin{tabular}{lccc}
Basis set&   N$_{\rm v}$ & {{$E^{\rm MP2}_{\rm ads}$}} &  {{$E^{\rm MP2(CP)}_{\rm ads}$}} \\ \hline
cc-pVDZ        &  328         & 510      & 157   \\
cc-pVTZ        &  762         & 365      & 195   \\
cc-pV(D,T)Z    &              & 304      & 211   \\
Canonical HFOs &   20523      & 217      & 217   \\
%RPA+SOSEX     & 228 & 216  \\
\end{tabular}
\end{ruledtabular}
\end{table}
%-------------------------------------------------------------------------

As a final application, we study water adsorption onto the surface of a lithium hydride crystal at the level of MP2 theory.
Although dissolution is the fate of this ionic crystal upon solvation, this process is first instigated by the adsorption of a single
water molecule, and the system has been studied extensively and with high accuracy by incremental methods and diffusion
Monte Carlo~\cite{Binnie2010,Binnie2011}.
In contrast to these, our work employs fully periodic boundary conditions, and the projected Gaussian space for the virtuals
is expected to be efficient in this cases since there is much vacuum required in the simulation cell to avoid spurious periodic images~\cite{Preuss04}.
Figure~\ref{fig:site} shows the relaxed structure of the adsorbed water molecule on the LiH
crystal. The structures have been relaxed using the DFT-PBE functional\cite{PBE}. Only the atoms
of the water molecule have been allowed to relax. The LiH surface
is modelled using a two layer surface supercell containing 16 Li and 16 H atoms. These atoms have been kept
fixed to the LiH crystal atom positions with a lattice constant corresponding to 4.1108~\AA.
The O 1$s$ states have been kept frozen in the MP2 calculation. All other electronic states have been treated
as valence states. %The first Brillouin zone has been sampled using the $\Gamma$-point only.

Table~\ref{tab:waterbinding2} summarizes the binding energies of the water molecule for 
different methods. The DFT-PBE functional yields a binding energy of 214~meV, which agrees well 
with the value of 212~meV reported in Ref.~\onlinecite{Binnie2011}.
On the level of Hartree--Fock, the water molecule exhibits a binding energy of 44~meV for the relaxed
structures. However, adding the electron correlation effects
on the level of MP2 theory yields
an adsorption energy for water of 157~meV and 195~meV for the pseudized cc-pVDZ and cc-pVTZ basis
sets, respectively. The MP2 calculations include CP corrections for the BSSE.
A simple CBS limit extrapolation yields an adsorption energy of 211~meV. %, very close to the MP2 result
%of 215~meV obtained using the incremental method reported in Ref.~\cite{Binnie2011}.

To further verify the PGTOs approach we have also performed a calculation with the full set of canonical Hartree--Fock orbitals
constructed from diagonalization of the Fock operator in the complete plane wave basis set. We note that this approach
additionally employs a basis set extrapolation technique which is outlined in Ref.~\onlinecite{Marsman2009}.
The obtained adsorption energy of 217~meV is in very good agreement with the other complete basis set limit findings within the far smaller PGTO basis sets, 
summarized in Table~\ref{tab:waterbinding3}. However, if this huge set of canonical virtual orbitals is truncated to similar sizes as the PGTO virtual space, 
then results are poor, and comparison to the bulk or isolated molecule is difficult, as it is hard
to truncate the canonical space consistently for the different systems.
%given the lack of physical truncation of the virtual space.
It is remarkable to see that PGTOs
and complete canonical HF orbitals yield results that agree to within 6~meV.
However, we stress again that it is extremely important to correct for BSSEs in the calculations as can be seen by comparing
$E^{\rm MP2}_{\rm ads}$ and $E^{\rm MP2(CP)}_{\rm ads}$ in Table~\ref{tab:waterbinding3}, where the BSSE is larger than the binding
energy itself.
Furthermore Table~\ref{tab:waterbinding3} also presents the number of virtual orbitals employed in the different calculations.
The biggest basis set in the PGTOs and the full plane wave basis in the canonical HFOs calculations corresponds to 762 and 20,523
orbitals, respectively. This comparison demonstrates strongly how much more compact the PGTOs basis for such systems can become
compared to canonical HF orbitals.
%We note that these findings are in good agreement with findings reported in
%Ref.~\cite{Binnie} using incremental methods. Ref.~\cite{Binnie} reports adsorption energies for the water
%molecule on the level of MP2 theory that are on the order of 210~meV. The achieved agreement is
%on the order of 30~meV and below chemical accuracy.
In a future study we will investigate this system in greater detail including
methods that also go beyond MP2 theory, as correlations beyond this level are clearly important, as can be seen by comparison to the CCSD(T) incremental
results, and diffusion Monte Carlo.

\section{Conclusion}
\label{sec:conclusion}

In this paper we have outlined a simple, black-box and robust approach to use local, atom-centered Gaussian basis functions within
a plane wave basis set using the projector augmented wave method and periodic boundary conditions.
The so-called pseudized Gaussian basis set can be expanded efficiently in a plane wave basis set with a moderate
kinetic energy cutoff. We have shown that a hybrid approach whereby the occupied orbitals are expanded in a
plane wave basis set and only the virtual orbital manifold is expanded in the pseudized Gaussian basis set orthogonalized
to the occupied orbitals provides a compact and systematically improvable basis.
The advantages of this over pure plane wave basis set calculations become most
beneficial in correlated wavefunction based calculations of low dimensional systems and systems where weak interactions
need to be described with high accuracy. This is not unexpected since the size of the plane wave basis set suffers from the
fact that it grows linearly with the box size regardless of the position or number of atoms in the studied system. As a result
it is difficult to devise systematic virtual orbital manifold truncation schemes for rapidly convergent energy differences.
In contrast to plane waves, Gaussian atom-centered basis sets take the local character of electronic
correlation into account {\it a priori} and allow for a systematic description of electronic correlation effects
such as the polarizability using a system tailored and compact basis set.

Of course, the introduction of such local basis sets
also always bears the burden of several shortcomings such as basis set superposition errors (BSSE) and linear dependencies
of diffuse atom-centered basis functions in densely packed solids.
These problems can partly be accounted for by counterpoise BSSE corrections and removing linearly dependent basis functions.
The compromise of a wavefunction expansion in Gaussians for the virtual space and a plane wave expansion for the occupied space
seems an efficient approach for combining the advantages and mitigating the disadvantages of each basis.
%Nonetheless it is clear that some of the intrinsic problems with GTOs also translate to the PGTOs described in this work.
We note that our method allows to easily switch between these two different basis sets (local atom-centered Gaussians and periodic
plane wave), which could
potentially lead to novel, transferable and more compact basis sets with the aim to reduce the computational cost of
correlated wavefunction based theories in periodic systems even further. In the future, this infrastructure will be combined with other
correlated methods, including coupled-cluster and F12 methods~\cite{Gruneis2011a,Gruneis2009,Booth2013,Gruneis2013a,PhysRevLett.115.066402,Usvyat2013}.
Furthermore, methods which directly exploit the locality of correlation effects, including quantum cluster methods such as dynamical mean-field theory~\cite{RevModPhys.78.865}
and density matrix embedding theory~\cite{PhysRevLett.109.186404,PhysRevB.93.035126,Scuseria_14,PhysRevB.91.155107}, as well as more traditional domain-based approaches to local correlation~\cite{LocalMP2,LocalMP2_2,Werner15} can be used within this framework, and
are being actively explored.

\section{Acknowledgments}
We gratefully acknowledge help with the implementation in \texttt{VASP} and useful discussions with Georg Kresse and Martijn Marsman. 
G.H.B gratefully acknowledges funding from the Royal Society. G. K.-L. Chan acknowledges support from the US Department
of Energy through grant DE-SC0010530, with secondary support from grant DE-SC0008624 (SciDAC).

\end{document}